\documentclass[usenatbib]{mnras}
\usepackage{graphicx}
\usepackage{color}
\usepackage{setspace}
\usepackage{dcolumn}
\newcolumntype{d}[1]{D{.}{.}{#1}}

\makeatletter
\newlength{\abovecaptionskip}%
\makeatother
\usepackage{threeparttable}

\usepackage{multicol}
\usepackage{pdflscape}

\def\apj{\rm ApJ}
\def\apjl{\rm ApJL}
\def\apjs{\rm ApJS}
\def\aj{\rm AJ}
\def\mnras{\rm MNRAS}
\def\nat{\rm Nature}

\def\aap{\rm AAP}

\def\gax{\mathrel{\raise.3ex\hbox{$>$}\mkern-14mu\lower0.6ex\hbox{$\sim$}}}
\def\lax{\mathrel{\raise.3ex\hbox{$<$}\mkern-14mu\lower0.6ex\hbox{$\sim$}}}
\def\gtorder{\mathrel{\raise.3ex\hbox{$>$}\mkern-14mu
             \lower0.6ex\hbox{$\sim$}}}
\def\ltorder{\mathrel{\raise.3ex\hbox{$<$}\mkern-14mu
             \lower0.6ex\hbox{$\sim$}}}

\def\dol{D_{OL}}
\def\dos{D_{OS}}
\def\dls{D_{LS}}

\begin{document}

\title [Variable Quasar Time Delays]
   {Microlensing Makes Lensed Quasar Time Delays Significantly Time Variable}

\author[Tie \& Kochanek]{ 
    S.~S. Tie$^1$,
    C.~S. Kochanek$^{1,2}$
    \\
  $^{1}$ Department of Astronomy, The Ohio State University, 140 West 18th Avenue, Columbus OH 43210 \\
  $^{2}$ Center for Cosmology and AstroParticle Physics, The Ohio State University,
    191 W. Woodruff Avenue, Columbus OH 43210 \\
   }

\maketitle
\begin{abstract}
The time delays of gravitationally lensed quasars are generally believed to be unique
numbers whose measurement is limited only by the quality of the light curves and the
models for the contaminating contribution of gravitational microlensing to the light curves.  
This belief is incorrect -- gravitational microlensing also produces changes in the
actual time delays on the $\sim$day(s) light-crossing time scale of the emission region. 
This is due to a combination of the inclination of the disk relative to the line of
sight and the differential magnification of the temperature fluctuations producing
the variability.  We demonstrate this both mathematically and with direct calculations
using microlensing magnification patterns. Measuring these delay fluctuations can provide
a physical scale for microlensing observations, removing the need for priors on 
either the microlens masses or the component velocities.  That time delays in lensed quasars are themselves
time variable likely explains why repeated delay measurements of individual lensed quasars
appear to vary by more than their estimated uncertainties.  This effect is also an
important new systematic problem for attempts to use time delays in lensed quasars for cosmology or to detect substructures (satellites) in lens galaxies.
\end{abstract}


\section{Introduction}

Ever since \cite{Refsdal1964} proposed the method, there have been hopes that 
gravitational lens time delays can be used to constrain the cosmological 
model (see the review by \citealt{Treu2016}).  Current studies are quite
optimistic, with \cite{Bonvin2017}, based also on \cite{Fassnacht2002},
\cite{Suyu2010}, \cite{Tewes2013b}, and \cite{Suyu2014},
 claiming a 3.8\% measurement of 
(effectively) $H_0$ based on time delay measurements in 3 lenses.  
Fundamentally, time delays constrain 
$\Delta t \propto H_0^{-1} (1-\langle \kappa \rangle)$
where $\langle \kappa \rangle$ is the mean surface density
(convergence) in the annulus between the two images 
(\citealt{Kochanek2002}). The surface density $\langle \kappa\rangle$
must be constrained with some additional information
about the geometry or kinematics of the lens (see the
review by \citealt{Kochanek2006}).  
One contribution to $\langle \kappa \rangle$
is simply the random fluctuations in the density along the line
of sight to the lens (see, e.g., \citealt{Keeton2004},
\citealt{Greene2013}, \citealt{McCully2017}), 
and the other is the internal structure of
the lens galaxy (see the discussion in \citealt{Kochanek2006}). 

The essence of the program to use lenses
to constrain cosmology is to obtain time delays of $N$ lenses
with individual time delay, line-of-sight convergence, and
lens surface density uncertainties of $\sigma_t$, $\sigma_{los}$
and $\sigma_{\langle\kappa\rangle}$ and then combine them to
obtain a fractional error in $H_0$ (really the combination of cosmological
distances entering the lens time delay) of 
$\sigma_{H}^2/H_0^2 \sim 
(\sigma_{\delta t}^2/\delta t^2+\sigma_{los}^2 + \sigma_{\langle \kappa \rangle}^2)/N$.  
It is generally assumed that the first two terms in this error
budget are dominated by random errors and their contributions
to the overall error will scale as $N^{-1/2}$.  There are 
far greater concerns about whether the last term may already
be dominated by systematic errors (e.g., \citealt{Schneider2013}, 
\citealt{Birrer2016}) for which there is no benefit from 
combining lenses.  In any case, any ultimate claim of achieving   
$\sigma_H/H_0 \simeq 0.01$ is equivalent to the claim that
there are also no systematic errors in the surface density
$\langle \kappa\rangle$ estimates at a comparable level.  It will be
challenging to prove this assertion.

This is not, however, the focus of the present paper.  Our concern 
here is the accuracy of the time delay measurements -- the error 
contribution from $\sigma_{\delta t}/\delta t$.  Dating from the
attempts to measure gravitational lens time delays in
the very first lens, Q~0957+561  (\citealt{Walsh1979}), time
delay measurements have produced controversy (e.g., \cite{Schild1990}
versus \cite{Press1992}, resolved in favor of the former by
\cite{Kundic1997}).  In most modern studies, the question is
not so much the basic validity of the delay but the accuracy
of the uncertainty estimates, and there have recently been
a series of tests and comparisons of various delay measurement
methods (e.g., \citealt{Tewes2013a}, \citealt{Dobler2015},
\citealt{Liao2015}, \citealt{Bonvin2017}).  An adequate 
summary of these studies is that with reasonably good light 
curves it is feasible to measure time delays with both high
accuracy and precision.

All of these studies assume, however, that the measured time delay
is the standard cosmological delay used in all lens models.
They also assume that the ratios of delays depend only on the
large scale potential of the lens galaxy (possibly with some
effects from the largest substructures in the galaxy, 
\citealt{Keeton2009}).  In this paper we show that for 
gravitationally lensed quasars, microlensing by the stars
in the lens galaxy (see the review by \citealt{Wambsganss2006})
makes both of these assumptions incorrect.

{\bf Microlensing changes time delays on the scale of the light
crossing time of the accretion disk, which has a typical scale of light days. These microlensing induced time
delays will then slowly change as the accretion disk moves 
relative the stars doing the microlensing. }

There are two causes of the microlensing effect on time delays.
The first is very simple.  If the accretion disk
does not lie in the plane of the sky, different parts of the disk
lie at different line of sight distances.  This has a negligible
consequence based on the normal expression for the time delay,
as partially discussed in \cite{Yonehara1999},
\cite{Goicoechea2002} and in more detail below.  However, the
normal time delay expression neglects the variations in the
delay due to the actual change in the line-of-sight distance $\Delta z$
to the source 
(because this term has been subtracted). This missing term
is a delay of order
$(1+z_s) \Delta z/c$ where $\Delta z/c$ is the proper light
travel time and $1+z_s$ is the effect of time dilation from the source redshift $z_{s}$.  Without microlensing, this effect is unimportant because $\langle \Delta z \rangle $ is the same for all images.
However, microlensing, by its very definition, differentially
magnifies different regions of the disk and
$\langle \Delta z \rangle$ will vary between images when weighted by the 
spatially varying microlensing magnification.  


The second effect is more subtle and exists even for a face-on
disk.  The variability of the disk
is due to a pattern of temperature fluctuations on the disk 
with the observed light curve representing an average of the
fluctuations.  The simplest case to explain is a ``lamp post''
model, where luminosity fluctuations $\delta L(t)$ close to the disk 
center illuminate the disk to drive the temperature fluctuations 
but with a lag due to the light travel time from the center,
$\delta T(R,t) \sim \delta L(t-R/c)$. While there is no guarantee that the lamp post model explains all quasar variability (see, e.g., \citealt{Dexter2011}), it has been successfully used to model the wavelength-dependent ultraviolet/optical/near-IR variability 
of NGC~2617 \citep{Shappee2014} and NGC~5548 \citep{Starkey2017}. The black body function is 
quite broad, so a broad range of disk radii contribute to
the variability observed at any given wavelength.  Microlensing
differentially weights the emission from the disk, so the
mean emission radius of a microlensed disk is different 
from that of the unmicrolensed disk or a differently microlensed
image of the same disk.  This both introduces differential time delays 
on the scale of the light crossing time, and means that the
light curves of lensed images will not be identical even for
a fixed microlensing magnification pattern.  There has
been some prior discussion of this latter point (e.g.,
\citealt{Gould1997}, \citealt{Wyithe2002}, \citealt{Dexter2011}), 
largely to explain the rapid, uncorrelated variability 
seen in the light curves of some lensed quasars (e.g., 
\citealt{Burud2002}, \citealt{Schechter2003}).
While we will illustrate this effect using the ``lamp post'' model,
the effect will be present for any model of quasar variability.

To summarize, we should expect gravitational lens time delays
to be affected by microlensing, with image-to-image shifts on
the time scale of the light crossing time of the disk -- days.
In \S2 we work through the basic mathematics of the effects.
In \S3 explore several numerical examples based on 
two of the lensed quasars used
in \cite{Bonvin2017}, RXJ~1131$-$1231 (\citealt{Sluse2003}) 
and HE~0435$-$1223 (\citealt{Wisotzki2002}). We summarize
the consequences of our results in \S4.  

\begin{figure}
\centering
\includegraphics[width=0.45\textwidth]{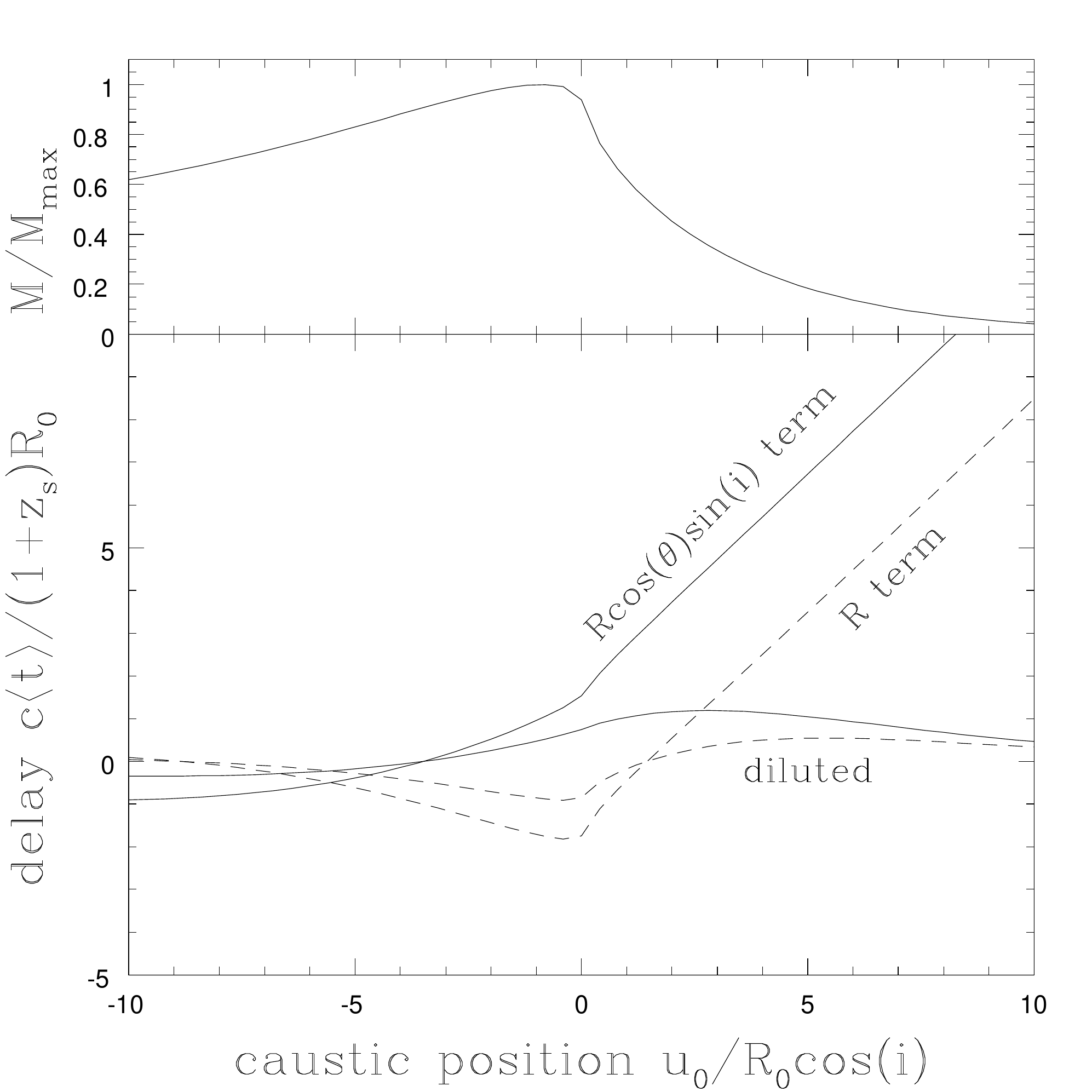}
\caption{ The observed mean time delay $\langle t \rangle$ in units of $(1+z_s)R_0/c$ (lower panel) and magnification
  relative to peak $M/M_{max}$ (top panel) for a linear fold caustic moving across a
  disk.  The caustic is oriented parallel to the long axis of the disk and
  its position $u_0/R_0 \cos i$ is in units of the projected, short axis of the 
  disk with the disk center at the origin. The solid line shows the inclination-dependent
  term ($R\cos\theta\sin i$) and the dashed line shows the inclination-independent term
  ($R$).  The lower amplitude, ``diluted'' curves show the effect of adding flux equal
  to that produced by the caustic at its peak and with no lag.
  } 
\label{fig:gauss}
\end{figure}

\section{Mathematics}

\def\bft{\vec \theta}
\def\bfb{\vec \beta}
The standard expression for the time delay of a lens is
\begin{equation}  
  \tau_{std} = 
   { \dol \dos \over c \dls } \left[ { 1\over 2 } \left(\bft-\bfb\right)^2 - \phi(\bft) \right]
    \label{eqn:std}
\end{equation}
where $\bft$ and $\bfb$ are the angular position of the image and the source and
$\phi(\bft) = (\dls/\dos) \phi_0(\bft)$ is the projected lensing potential (e.g., \citealt{Schneider1985}, \citealt{Blandford1986}). The
distances are proper motion distances between the Observer, Lens and Source.
For a flat universe they are simply the comoving distances and 
$\dos=\dol+\dls$.  Using these
distances instead of angular diameter distances both allows the use of this simple relation and eliminates extra redshift factors\footnote{This is true of most lensing calculations, see \cite{Kochanek1993}.}.  The
lens potential combines a function $\phi_0(\bft)$ which is independent of the 
source distance with a source distance scaling of $\dls/\dos$ for all models.  Using Fermat's
principle, the observed images are located at solutions of 
${\vec \nabla}_{\bft} \tau_{std}=0$.

Now consider a second source displaced in angle to $\bfb+\delta\bfb$
and source redshift to $z_s + \delta z_s$.  The images of this source
will have different time delays due to the shift in the source position
and the change in the Einstein radius of the lens.  If we define
$F=D_{OS}/D_{LS}$, the change in the delay for any two associated
images is
\begin{equation}
  \Delta \tau_{std} =
    { \dol \dos \over c \dls } \left[ \delta\bfb \cdot \left( \bfb - \bft \right)  
     + { 1 \over 2 } { \delta z_s \over F } { d F \over dz_s }
              \left( \bfb - \bft \right)^2 \right],
\end{equation}
where the first term is the one previously discussed by \cite{Yonehara1999}
and \cite{Goicoechea2002}.  The overall delays have a general scale of
$(\dol \dos /c \dls ) \Delta\theta^2$, where $\Delta\theta^2$
is the image separation (see, e.g., the review by \citealt{Kochanek2006}). 
This means that the fractional change in the time delay from the first 
term is of order $|\delta \beta/\Delta \theta| \sim 10^{-5}$ for a disk
scale of order 10~light days and a lens Einstein radius of order a $1$~kpc.
Hence, as fully realized by \cite{Yonehara1999} and \cite{Goicoechea2002},
this effect only matters for sources separated by distances that are a
non-trivial fraction of $1$~kpc.  The second term was not considered in
these papers and produces a fractional correction of order 
$|(\delta z_s/F)(dF/dz_s)|$.  For a flat universe where $D = (c/H_0) \int dz/E(z)$
with $E(z) = H(z)/H_0$, this becomes $(c \delta z_s /H(z_s) \dos)(\dol/\dls)$
where the first term is essentially the fractional change in the distance
to the source, which is $\sim 10^{-12}$ and so even less important than the
first term.  At this point, the reader may be wondering why this paper is being
written. 

We have, however, left out an important effect. Equation~\ref{eqn:std}
is the time delay relative to a fiducial ray from the observer to the source
along the optic axis.  The actual total delay is not $\tau_{std}$ but
(for a flat universe)
\begin{equation}
        \tau = { D_{OS} \over c } + \tau_{std}.
\end{equation}
For the time delay between two sources at a common distance, the first term
simply cancels and thus is simply removed as a nuisance parameter to
derive Equation~\ref{eqn:std}.  For
two sources at different distances, however, it produces a delay difference
of $ (1+z_s) \Delta z /c$ where $\Delta z$ is the proper separation of the two
sources at the source and the net delay is simply the lag in the rest 
frame of the source time-dilated by the source redshift.  Unlike the two terms
that result from shifting the source in Equation~\ref{eqn:std}, this term
has the scale of the source size and it will matter for time delays between
two images if the light contributing to the first image has an average source 
distance different from that contributing to the second image.  

To calculate the delay shifts created by the inclination of the disk, we
require a disk model.  Since we also want to compute the second effect
created by differential microlensing of the variable flux, it should
also allow for a simple model of the variable flux.  We will consider 
a standard, non-relativistic, thin disk model emitting as a black body
(\citealt{Shakura1973}). If we are observing at wavelength rest $\lambda$,
it is useful to define
\begin{equation}
  \xi = { h c \over k T_0(R) \lambda }
      = \left( { R \over R_0}\right)^{3/4} 
        \left(1-\sqrt{ R_{in} \over R}\right)^{-1/4}
\end{equation}
where $R > R_{in}$, $T_0(R)^4\propto R^{-3}(1-\sqrt{R_{in}/R})$ is the temperature profile
of the disk,
\begin{eqnarray}
	\label{eqn:rdisk}
   R_0 &= &\left[ { 45 G \lambda_{rest}^4 M_{BH} \dot{M} \over 16 \pi^6 h_p c^2 } \right]^{1/3} \\
       &= &9.7 \times 10^{15} \left( { \lambda_{rest} \over \mu\hbox{m} }\right)^{4/3}
                  \left( { M_{BH} \over 10^9 M_\odot } \right)^{2/3} 
                  \left( { L \over \eta L_E } \right)^{1/3}~\hbox{cm} \nonumber
\end{eqnarray}
is the radius where the disk temperature matches the photon wavelength, $ k T = h_p c/\lambda_{rest}$,
and $R_{in}=\alpha GM_{BH}/c^2$ is the inner edge of the disk (e.g., \citealt{Morgan2010}). 
Here $h_p$ is the Planck constant, $k$ is the Boltzmann constant, $M_{BH}$ is the black hole mass,
$\dot{M}$ is the mass accretion rate, $L/L_E$ is luminosity in units of the Eddington
luminosity, $\eta = L/\dot{M}c^2$ is the accretion efficiency, and $\alpha = 6$ for a Schwarzschild black hole and $\alpha = 1$ for an equatorial orbit co-rotating Kerr black hole. Converted to a time scale and using the observed wavelength instead of the rest wavelength,
\begin{equation}
   { (1 + z_s) R_0 \over c }
     \simeq { 3.8~\hbox{days} \over  (1+z_s)^{1/3}} 
         \left( { \lambda_{obs} \over \mu\hbox{m} }\right)^{4/3}
                  \left( { M_{BH} \over 10^9 M_\odot } \right)^{2/3} 
                  \left( { L \over \eta L_E } \right)^{1/3}.
\end{equation}
The unperturbed surface brightness profile of the disk is simply
\begin{equation}
   I_0(R)\propto \left[ \exp(\xi) -1 \right]^{-1}.
\end{equation}
Ignoring the inner edge of the disk ($R_{in}\rightarrow 0$), the mean
radius of the unperturbed surface brightness profile is 
$\langle R I_0\rangle/\langle I_0 \rangle =  3.36 R_0$.
That we are using a monochromatic wavelength is unimportant, as the
radial width due to the black body emission is significantly more 
important than the wavelength spread from a typical broad-band filter.

In the simplest ``lamp post'' model of variability (e.g., \citealt{Cackett2007}),  the
fractional temperature variation is independent of radius in the disk, so
\begin{equation}
     T(R,t) = T_0(R) \left[ 1 + f(t-R/c) \right].
\end{equation} 
where $T_0(R)$ is the unperturbed temperature profile and $f(t-R/c)$ is
the fractional luminosity variability ``lagged'' by the light travel time $R/c$ from
the disk center.  If we assume the temperature variations are small,
then we can Taylor expand the black body function to find that the 
time-variable emission is
\begin{equation}
   \delta I(R,t) \propto  f(t-R/c) G(\xi) \quad\hbox{where}\quad 
   G(\xi) = { \xi \exp(\xi) \over \left(\exp(\xi)-1\right)^2}
   \label{eqn:gxi}
\end{equation}
comes from the temperature derivative of the black body and the
definition of $f$.  The average radius of the variable flux, 
$\langle R G \rangle/\langle G \rangle =  5.04 R_0$ is larger
than that of the unperturbed disk because a constant fractional 
temperature fluctuation produces larger surface brightness
fluctuations at larger radii where the disk becomes cooler. 

We can characterize the effects of microlensing by averaging the delays over 
the variable surface brightness $G(\xi)$ weighted by the absolute
value of the microlensing 
magnification $M(u,v)$. Let the disk be tilted relative to
the line of sight by an inclination angle $i$ with $i=0$ corresponding to
the disk lying in the plane of the sky.  If a point in the disk is labeled
by $(x,y,z)=R(\cos\theta,\sin\theta,0)$ and we rotate about the $y$ axis, then 
the observed position is $(u,v,w)=R(\cos\theta\cos i, \sin\theta, \cos\theta\sin i)$.
The average delay between the driving source $f(t)$ and the observer is
\begin{equation}
\langle  \delta t \rangle = { 1+z_s \over c } 
    { \int dudv G(\xi) M(u,v) R \left( 1 + \cos\theta \sin i \right)
       \over \int du dv G(\xi) M(u,v) }
   \label{eqn:anal}
\end{equation}
where $u= R \cos\theta \cos i$ and $v= R \sin\theta$ are coordinates on the
reference source plane of the lens, and $R^2=u^2/\cos^2 i + v^2$.
The factor governing the mean delay $R\left( 1 - \cos\theta \sin i \right)/c$
combines the propagation delay $R/c$ for the lamp post, with the 
line-of-sight delay $(R/c) \cos\theta \sin i$ due to the inclination
of the disk.  Without microlensing ($M(u,v) \equiv 1$) and ignoring the 
inner disk edge, the inclination of the disk has no effects and
we just measure the mean lag between the driving perturbation and
the observed light curve, $\langle \delta t \rangle = 5.04(1+z_s)R_0/c$.    

It is relatively easy to show that the primary lens potential and
its satellites cannot produce large enough magnification gradients
across the accretion disk to produce an observable effect.
This is not true of microlensing of the disk by the
stars near each lensed image.   The basic physics of the effect,
that the star field near each image produces a different and 
time variable magnification of the accretion disk, means that
there are gradients in the magnification on the physical scale
of the disk and that the characteristic amplitude of the microlensing-induced
delays must be of order the light crossing time of the disk. 

Figure~\ref{fig:gauss} shows the two contributions to the mean
delay in Equation~\ref{eqn:anal} for a linear fold caustic 
parallel to the $v$ axis as a function of its position $u_0$
as it is moved across the disk.  The magnification produced
by the fold caustic is $M = A |u-u_0|^{-1/2}$ for $u\geq u_0$ 
and $0$ for $u < u_0$.  For the $R/c$ term of the delay, we
have subtracted the lag in the absence of microlensing in 
order to focus on the changes created by the effect.  Since
the fold is moving along the projected short axis of the disk,
the fold position is scaled by the projected short axis disk scale length,
$u_0/R_0\cos i$.  The scale of the delays is $(1+z_s)R_0/c$,
with the term due to the lag in the lamp post model ($R/c$)
being independent of the inclination, and the term due to
the disk inclination ($R\cos\theta\sin i$) depending on the
inclination.  We have set $R_{in}=R_0/100$, although the 
results are insensitive to this choice for the parameter
ranges relevant to optical monitoring of lensed quasars.
For our definitions, the near side of the disk has $u<0$ 
and the far side has $u>0$.  

We can qualitatively understand the behavior of both terms.
Consider the term due to the inclination of the disk first.
When $u_0$ is negative, the caustic lies on the near side
of the disk and so magnifies parts of the disk with shorter,
negative delays.  However, the fold magnification is not
very singular, and so at large distances the dropping flux
of the disk matters more than the magnification and the 
average delay becomes that of an unmicrolensed disk.  As
the caustic approaches the disk center, the magnification of 
the near side makes the lag increasingly negative, but 
not by huge factors.  The sign reverses before the caustic
reaches the disk center because the total flux from the
far side now exceeds that from the near side. Once the caustic
passes the disk center, the lag simply increases because
there is no longer any contribution from the near side of
the disk or smaller radii than the distance of the 
caustic from the disk center.

The term due to the lamp post model has some qualitative
differences.  It is again negligible when the caustic
is on the near side and at large radius, but slightly
positive because it is enhancing the contribution
from large radii.  As the caustic approaches the disk
center, it produces a negative mean lag by enhancing the
contribution from small radii.  Then, as it moves outwards,
it also produces an increasingly positive lag for the same
reasons as the inclination-dependent term.  

The flux from a microlensed quasar is never due to a single fold
caustic.  It is the sum of the flux from the direct image plus
some number of additional image pairs created by microlensing
(see, e.g., \citealt{Granot2003}).  The
model with a single fold caustic explores the contribution
of a single image pair created by a fold caustic, which
must then be combined with the fluxes of all the other 
images.  When the flux from this image pair is small 
compared to the flux of the other images, the observed 
delay will be dominated the delays associated with the
other images. Figure~\ref{fig:gauss}
also shows the magnification produced by the caustic
relative to its peak, $M/M_{max}$.  The magnification
rises until the caustic approaches the disk center and
then declines.  Thus, when the caustic is producing the
largest shift in the time delay, it is also making a
smaller and smaller contribution to the observed flux
because only the outer parts of the disk are being lensed by the fold caustic.

We can get a sense of the consequences by assuming that
there is additional flux equal to the flux produced by
the caustic at its peak and with an average lag of zero.
Under these assumptions, the mean lag becomes 
$\langle \delta t \rangle M/(M+M_{max})$, which is also shown in
Figure~\ref{fig:gauss}.  The dilution by the additional
flux eliminates the divergence in the delay as the 
caustic moves out to increasingly large radii. 
Reality is more complex, which is why numerical simulations
are ultimately required to understand the magnitude of the
effect.

\section{Numerical Simulations}
 We demonstrate the effects of microlensing on gravitational lens time delays using numerical simulations for two of the lensed quasars, RXJ~1131$-$1231 and HE~0435$-$1223, used by \cite{Suyu2014} and \cite{Bonvin2017} to constrain $H_{0}$. RXJ~1131$-$1231 is a four image lensed quasar with $z_{s}$ = 0.658 and $z_{l}$ = 0.295 \citep{Sluse2003}. Based on the H$\beta$ line width from \cite{Sluse2003}, \cite{Dai2010} estimated the black hole mass to be (1.3 $\pm$ 0.3)$\times$10$^{8}$ $M_{\odot}$. HE~0435$-$1223 is another four image lensed quasar with $z_{s}$ = 1.689, $z_{l}$ = 0.46, and an estimated black hole mass of 0.5$\times$10$^{9}$ $M_{\odot}$ (see \citealp{Mosquera2011} and references therein). Based on Equation 5, the accretion disk sizes for RXJ~1131$-$1231 and HE~0435$-$1223 are $R_0 = 7.34 \times 10^{14}$ cm (0.28 light days) and $R_0 = 9.37 \times 10^{14}$ cm (0.36 light days), respectively, in the observed $R$-band (6586 \AA) for an Eddington ratio of $L/L_{E}$ = 0.1 and a radiative efficiency of $\eta$ = 0.1. 

\begin{figure*}
\begin{multicols}{2}
    \includegraphics[width=0.45\textwidth]{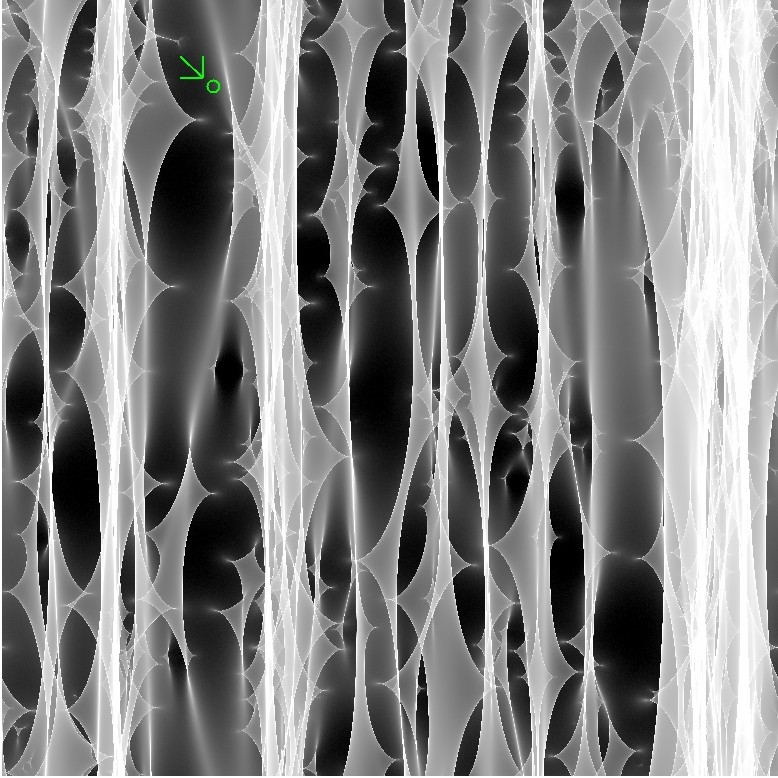}
    \caption{Magnification pattern for image A of RXJ~1131$-$1231. The magnification increases from darker/de-magnified to brighter/magnified regions. The green circle on the upper left shows the mean radius of the variability-dominated region, $\sim$5$R_0$. The pattern spans 20$\langle R_{e} \rangle$.} 
	\label{fig:map1131}
    
    \includegraphics[width=0.45\textwidth]{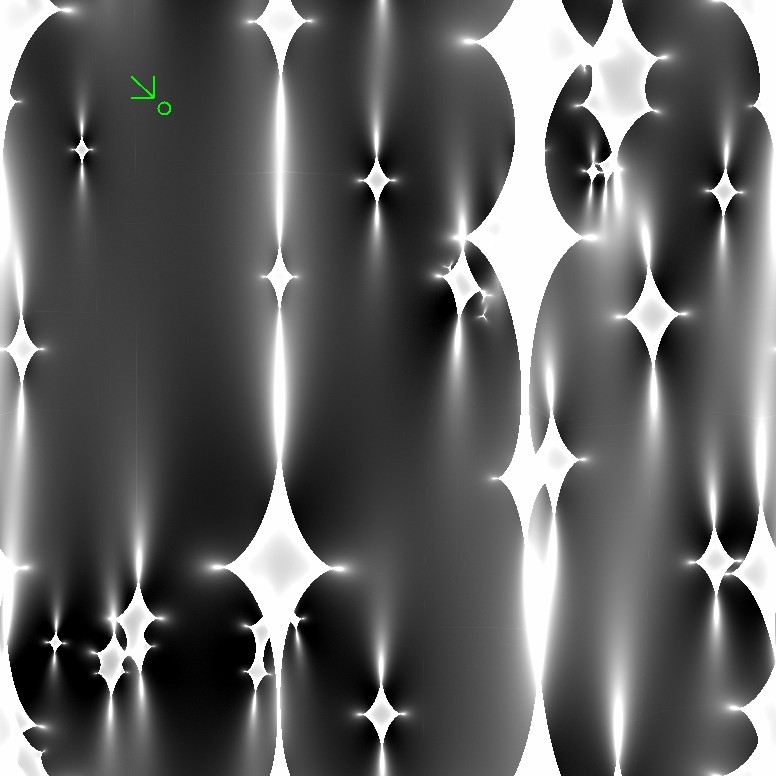}
	\caption{Magnification pattern for image A of HE~0435$-$1223. The format is the same as for Figure~\ref{fig:map1131}.}
	\label{fig:map0435}
\end{multicols}    
\begin{multicols}{2}
    \includegraphics[width=0.45\textwidth]{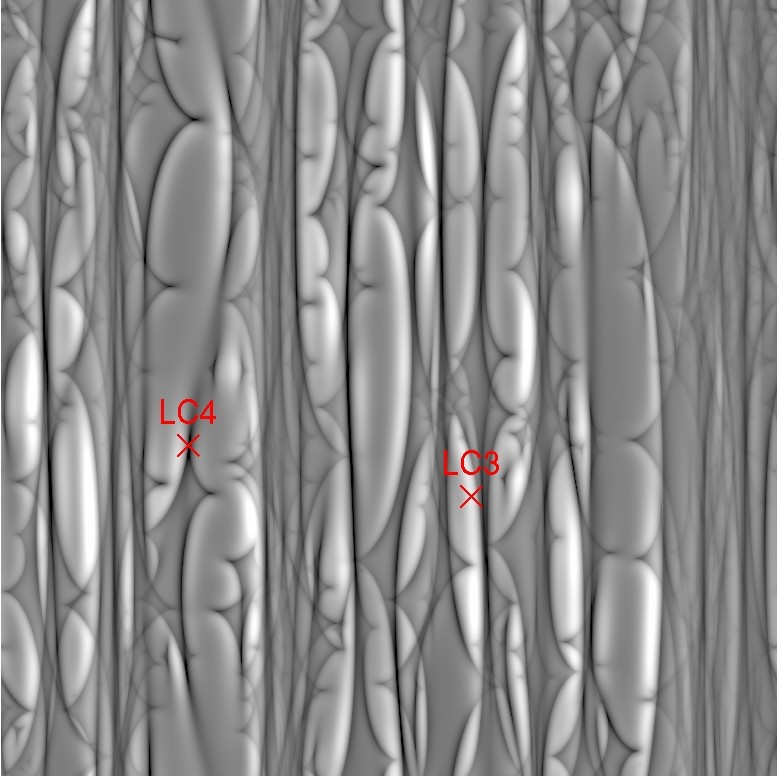}
	\caption{Delay map for image A of RXJ~1131$-$1231, where the disk is inclined to $i=$ 30$^{\circ}$  and rotated to $PA=$ 45$^{\circ}$ for the source size $R_0$. The mean delays span from $-1.35$ days to +3.91 days and  increase from the dark to the bright regions, with black indicating negative delays. The red crosses mark the points used for the light curves LC3 and LC4 in Figure \ref{fig:exmaplelc}.} 
	\label{fig:delay1131}
    
    \includegraphics[width=0.45\textwidth]{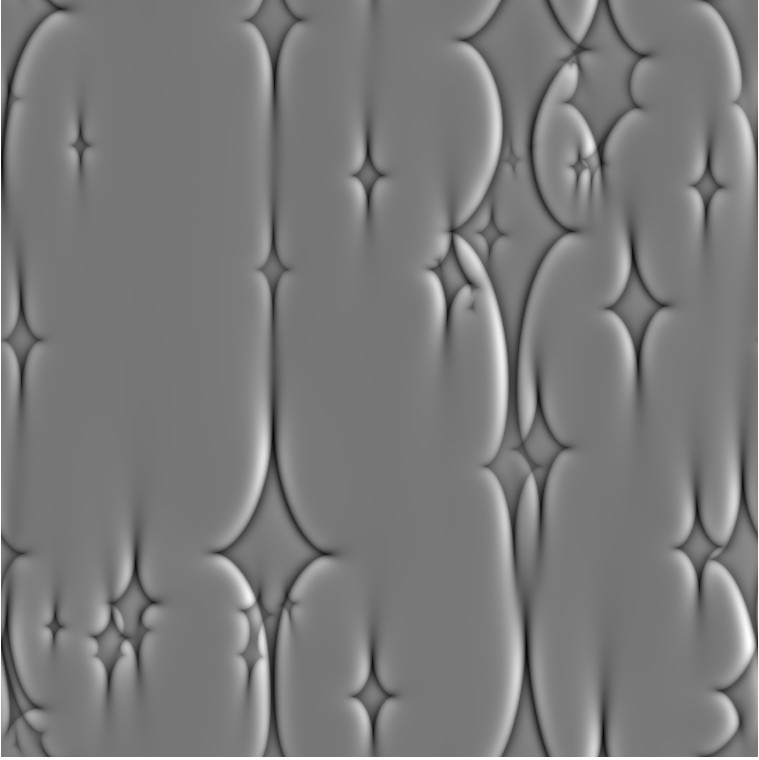}
	\caption{Delay map for image A of HE~0435$-$1223 using the same disk configuration as in Figure \ref{fig:delay1131}. The mean delays span $-2.27$ days to +2.59 days and increase from dark to bright regions, with negative delays shown in black. } 
	\label{fig:delay0435}
\end{multicols}
\end{figure*}

We created magnification maps for each lensed image using the ray-shooting method described in \cite{Kochanek2004}. The microlensing parameters, given in Table \ref{tab:microparam}, correspond to a macro model with a stellar mass fraction of 0.2 relative to a pure de Vaucouleurs model \citep{Dai2010}. The magnification maps have dimensions of 8192$\times$8192, an outer scale of 20$\langle R_{e} \rangle$, and a pixel scale of 0.00244 $\langle R_{e} \rangle$, where $\langle R_{e} \rangle$ is the Einstein radius at the source plane. We assume a mean microlens mass of $\langle M/M_{\odot} \rangle = 0.3$ for both lenses, leading to an outer scale of $5.02 \times 10^{17}$ cm and a pixel scale of $6.12 \times 10^{13}$ cm for RXJ~1131$-$1231 and an outer scale of $5.89 \times 10^{17}$ cm and a pixel scale of $7.19 \times 10^{13}$ cm for HE~0435$-$1223. The accretion disk scale lengths $R_0$  are $\sim$ 12 times the pixel scale, while the region dominating the variability ($\sim$ 5$R_{0}$) is $\sim$ 60 times the pixel scale. Figure \ref{fig:map1131} and \ref{fig:map0435} show the full 8192$\times$8192 magnification patterns for image A of each lens.

\begin{table}
\small
\caption{Microlensing model parameters} 
\centering 
\begin{tabular}{c c c c c}
\hline\hline 
Lens & Image & $\kappa$ & $\gamma$ & $\kappa_{*}/\kappa$ \\
\hline
RXJ~1131$-$1231 & A & 0.618 & 0.412 & 0.0667\\
                & B & 0.581 & 0.367 & 0.0597\\
                & C & 0.595 & 0.346 & 0.0622\\
                & D & 1.041 & 0.631 & 0.1590\\[1.0ex]
                
HE~0435$-$1223  & A & 0.604 & 0.262 & 0.0500 \\
                & B & 0.734 & 0.395 & 0.0801 \\
                & C & 0.605 & 0.265 & 0.0500 \\
                & D & 0.783 & 0.427 & 0.0930\\

\hline
\end{tabular}
\label{tab:microparam}
\end{table}

We calculated the mean delays between the driving source (the ``lamp post'') and the observer using Equation~\ref{eqn:anal} for each lensed image, where the surface brightness $G(\xi)$ is given by Equation~\ref{eqn:gxi} and the inner edge of the disk is (safely) ignored. We investigated four disk configurations with inclination $i$ and position $PA$ angles of (i) $i$ = 30$^{\circ}$, $PA$ = 0$^{\circ}$, (ii) $i$ = 30$^{\circ}$, $PA$ = 45$^{\circ}$, (iii) $i$ = 30$^{\circ}$, $PA$ = 90$^{\circ}$, and (iv) $i$ = 0$^{\circ}$. The position angles determine whether the long axis of the tilted disk is parallel, perpendicular, or at an angle to the caustic structures in the magnification maps. In this paper, the long axis of the tilted disk is perpendicular (parallel) to the caustic structures for $PA$ = 0$^{\circ}$ ($PA$ = 90$^{\circ}$). Note that the last case (iv) corresponds to a face-on disk where the position angle does not matter. For each disk configuration, we also investigated the effect of decreasing and increasing the source size by a factor of two. 

Examples of mean delay maps for image A of the two lenses are shown in Figures \ref{fig:delay1131} and \ref{fig:delay0435}, where the constant delay $\langle \delta t \rangle = 5.04(1+z_s)R_0/c$ of the lamp post model has been subtracted. The maps represent time delay perturbations due to microlensing, typically on the order of $\sim$ a day, with both positive and negative delays. The negative delays are shown in black, while the positive delays increase from darker to brighter regions. The edges and structures where the time delays change rapidly and frequently flip signs are due to caustics in the corresponding magnification patterns, shown in Figures~\ref{fig:map1131} and \ref{fig:map0435}. 

\begin{figure*}
\centering
	\includegraphics[width=0.97\textwidth]{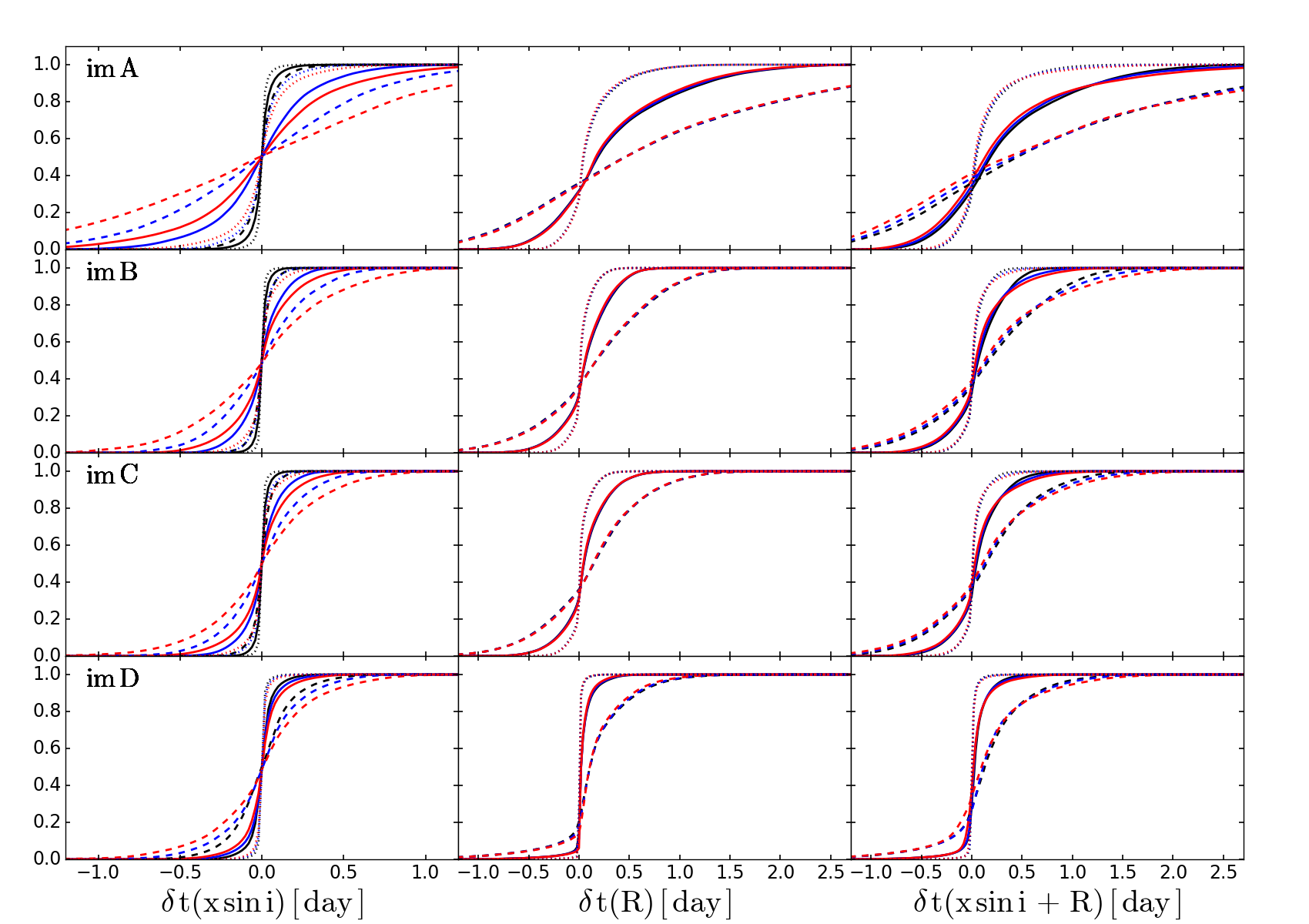}
	\caption{Cumulative distributions of the microlensing-induced mean delays for RXJ~1131$-$1231 when the disk is inclined to $i=$ 30$^{\circ}$. The row refers to the different lensed images while the column refers to the different delay contributions: from left to right, the delay component from the LOS inclination term, the $R/c$ term, and their combination. The different colors indicate different disk position angles: black for $PA$ = 0$^{\circ}$, blue for $PA$ = 45$^{\circ}$, and red for $PA$ = 90$^{\circ}$. The different line styles refer to the different source sizes:  $R_{0}$ as the solid line, $2R_{0}$ as the dashed line, and 0.5$R_{0}$ as the dotted line.} 
	\label{fig:hist1131}
\end{figure*}

\begin{figure*}
\centering
\includegraphics[width=0.97\textwidth]{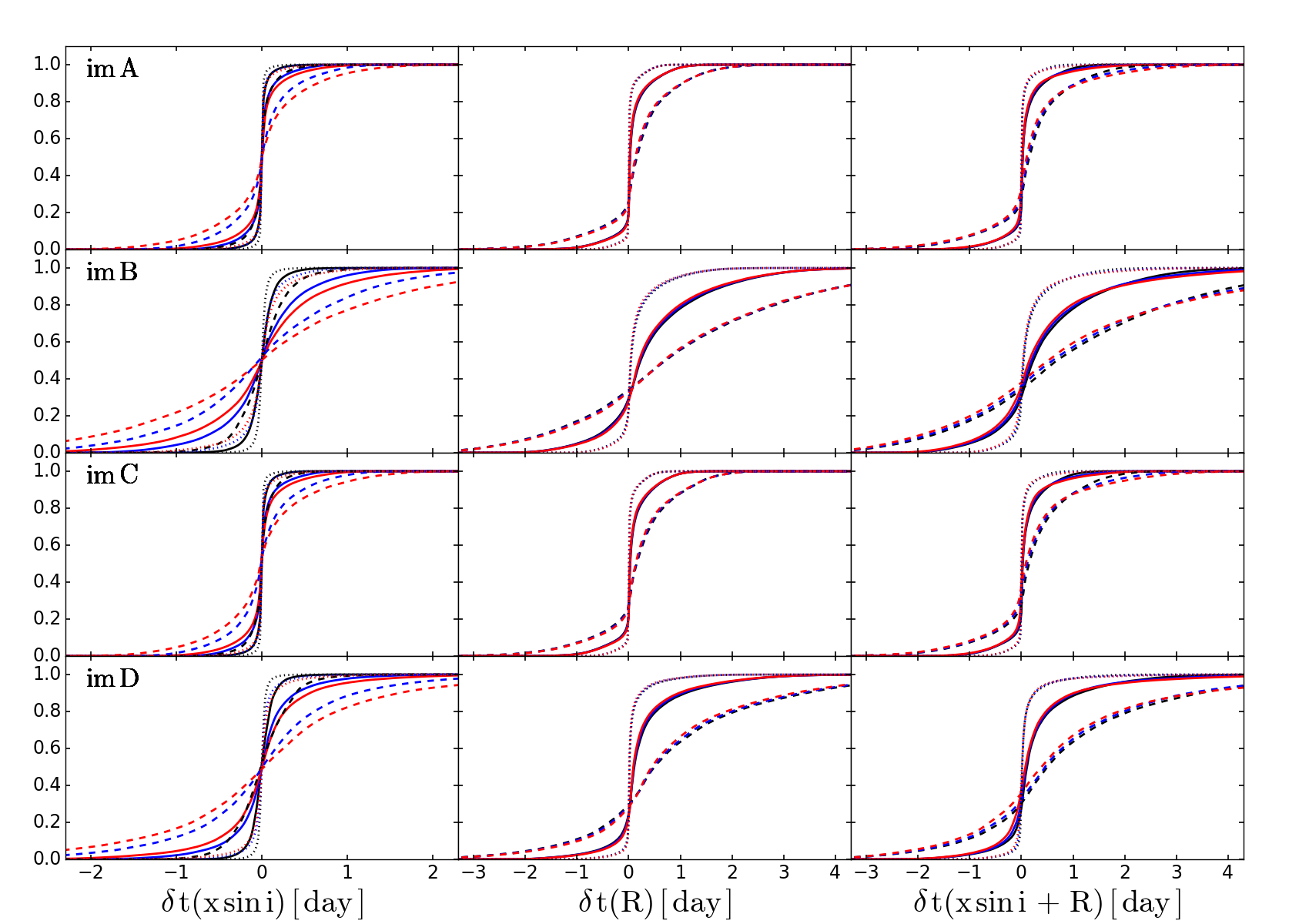}
\caption{Same as Figure \ref{fig:hist1131}, but for HE~0435$-$1223.} 
\label{fig:hist0435}
\end{figure*}

\begin{figure*}
\centering
\includegraphics[width=0.97\textwidth]{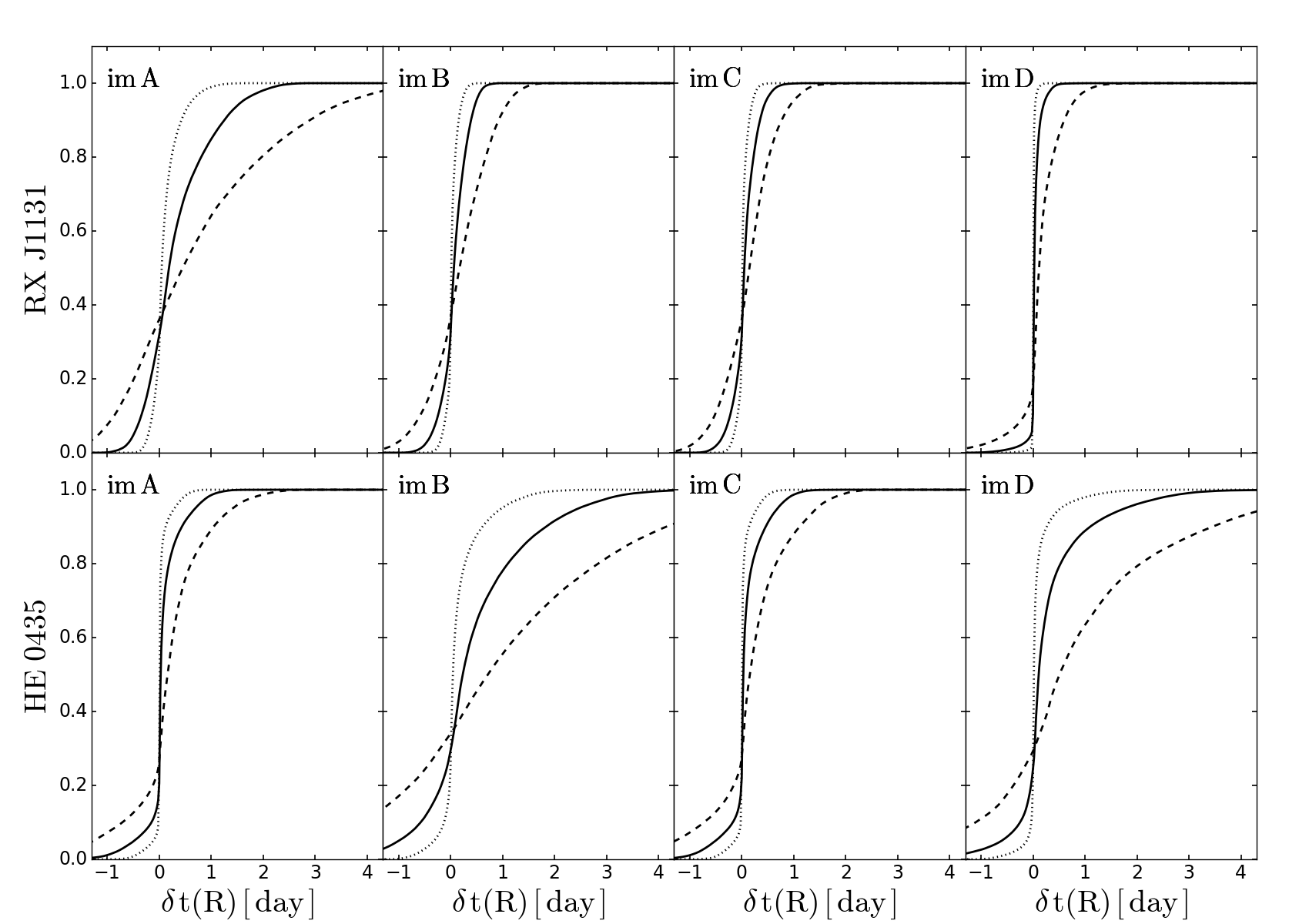}
\caption{Cumulative distributions of the mean delays for a face-on disk. The top row is for RXJ~1131$-$1231 and the bottom row is for HE~0435$-$1223. The columns refer to the four lensed images. The disk $PA$ does not matter for a face-on disk. The different line styles refer to source sizes of $0.5R_0$ (dotted), $R_0$ (solid) and $2R_0$ (dashed).} 
\label{fig:histfaceon}
\end{figure*}

Figures \ref{fig:hist1131}$-$\ref{fig:histfaceon} show the mean delay distributions for the different disk configurations for all images of the lenses, based on $\sim$ 300,000 randomly-selected points from their delay maps. There are several interesting points to note from the delay distributions. First, the line of sight (LOS) delays from the disk inclination have a zero mean and are symmetric about zero delay, as expected. On the other hand, the $R/c$ delays depend little on inclination and have a positive mean. The skew to positive delays can be understood as follows. Producing a negative lag for the $R/c$ term requires magnifying only the inner part of the disk, which rarely happens because the caustic structures generally have scales larger than $\sim$ 5$R_{0}$ and will magnify the outer parts of the disk as well. On the other hand, it is relatively easy to magnify the outer parts of the disk without magnifying the inner parts, leading to the delay asymmetry. Each image therefore has a non-zero mean for the total delay. 

\begin{figure*}
\centering
\includegraphics[width=0.97\textwidth]{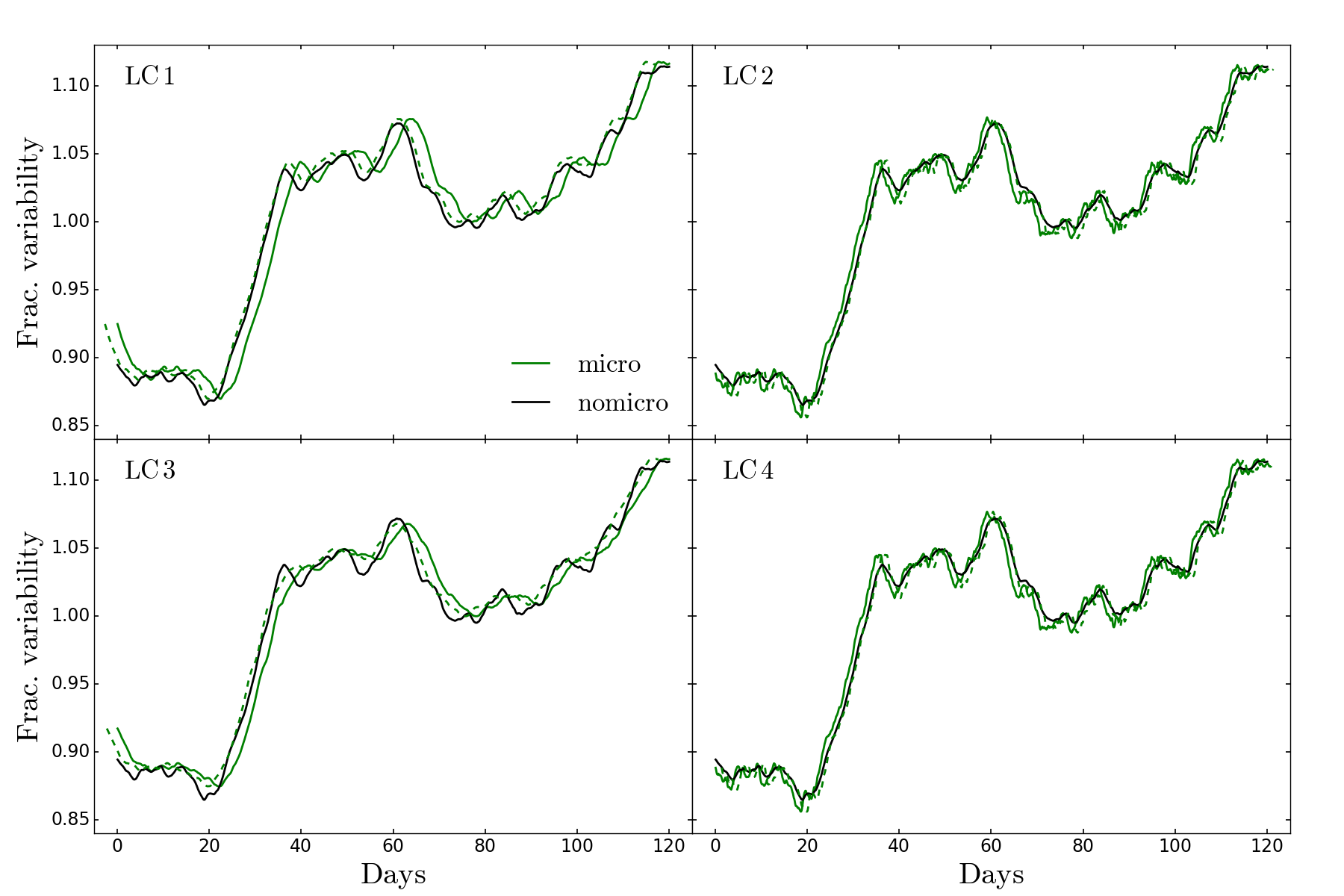}
\caption{Example light curves for image A of RXJ~1131$-$1231 at randomly-selected points in the map. The top panel shows two light curves for a face-on disk, while the bottom panel shows two light curves for the $i$ = 30$^{\circ}$ inclined disk at $PA$ = 45$^{\circ}$. The locations used for the inclined disk models are marked as ``LC3''  and ``LC4'' in Figure \ref{fig:delay1131}. The ``nomicro'' light curves in black correspond to no microlensing, while the ``micro'' light curves in green include the effects of microlensing. The dashed lines show the effect of removing the predicted mean delays (Table \ref{tab:javelin_lag}). Even with the mean delays removed, the light curves are not identical because microlensing weights the temperature variation of the disk differently than a uniform weighting. The ``micro'' light curves are ahead of the ``nomicro'' light curves (i.e. observed at later times) when the lag is positive and vice versa. } 
\label{fig:exmaplelc}
\end{figure*}

For all source sizes, the magnitudes of the total delays increase from a $PA$ of 0$^{\circ}$ (perpendicular to caustic structures) to a $PA$ of 90$^{\circ}$ (parallel to caustic structures). This is because when the long axis of the disk is perpendicular to the caustic structures, regions of positive and negative LOS delays are being magnified at the same time, leading to a smaller overall effect. When the caustic structures are parallel to the long axis of the disk, it is easier to magnify regions with only one sign of the delay. Therefore, microlensing has a more significant effect when the long axis of the disk is parallel to the caustic structures. Like the disk inclination, the disk $PA$ has a larger effect on the LOS delays than the $R/c$ delays. The overall delays are also larger for a larger source size and smaller for a smaller source. Even when the disk is face-on, there are still non-negligible microlensing delay contributions due to the light travel time from the lamp post to the disk. In all cases, we have subtracted the mean delay $\langle \delta t \rangle = 5.04(1+z_s)R_0/c$ of the lamp post model and verified that a uniform magnification leads to no change in the mean delays. 

Table~\ref{tab:rmsdelay} summarizes the mean and dispersion of the delay differences between images based on randomly drawing a large number of delays from the delay distributions of the respective images. Since the mean total delay for the individual images are non-zero, the mean of the delay differences is also not zero. These non-zero means represent a bias in the delays between the images that cannot be removed simply by monitoring the lens for a long period of time. The dispersion of the delay differences about this mean represents an additional scatter introduced into a time delay between two images, which can be eliminated by monitoring the lens of a long time period. In producing Table~\ref{tab:rmsdelay}, we simply used the same disk geometry for each image. In reality, the $PA$ of the disk relative to the magnification pattern of the other images is determined once it it set for the first image to the extent that the orientation of the model shear for each image is well constrained. 

\begin{table*}
\small
\caption{The mean (dispersion) of the microlensing time delay differences between image pairs}
\centering 
\begin{tabular}{cccccccccc}
\hline\hline 
Lens & Size & $PA$ $(^{\circ})$ & $i$ $(^{\circ})$ & $B-A$ & $C-A$ & $D-A$ & $C-B$ & $D-B$ & $D-C$ \\
 & & & & (days) & (days) & (days) & (days) & (days) & (days) \\
\hline
\rule{0pt}{2.5ex}
RXJ~1131$-$1231 & 0.5$R_0$ & 0 & 0 & $-0.08$ (0.27) & $-0.08$ (0.27) & $-0.10$ (0.25) & $-0.00$ (0.15) & $-0.02$ (0.11) & $-0.02$ (0.11)\\
                &  ... & 0 & 30 & $-0.08$ (0.27) & $-0.09$ (0.27) & $-0.10$ (0.26) & $-0.00$ (0.15) & $-0.02$ (0.12) & $-0.02$ (0.11)\\
                &  ... & 45 & 30 & $-0.08$ (0.30) & $-0.08$ (0.30) & $-0.10$ (0.28) & $-0.00$ (0.17) & $-0.02$ (0.13) & $-0.02$ (0.12)\\
                &  ... & 90 & 30 & $-0.07$ (0.34) & $-0.08$ (0.33) & $-0.09$ (0.31) & $-0.00$ (0.18) & $-0.02$ (0.14) & $-0.02$ (0.13)\\
                \rule{0pt}{3.5ex}
                & $R_0$ & 0 & 0 & $-0.25$ (0.68) & $-0.27$ (0.67) & $-0.29$ (0.64) & $-0.01$ (0.34) & $-0.04$ (0.28) & $-0.02$ (0.27)\\
                & ... & 0 & 30 & $-0.25$ (0.68) & $-0.27$ (0.68) & $-0.29$ (0.65) & $-0.01$ (0.35) & $-0.04$ (0.29) & $-0.02$ (0.28)\\
                & ... & 45 & 30 & $-0.24$ (0.74) & $-0.26$ (0.74) & $-0.28$ (0.71) & $-0.01$ (0.38) & $-0.04$ (0.32) & $-0.02$ (0.30)\\
                & ... & 90 & 30 & $-0.24$ (0.83) & $-0.25$ (0.81) & $-0.27$ (0.78) & $-0.01$ (0.42) & $-0.04$ (0.35) & $-0.02$ (0.33)\\
                \rule{0pt}{3.5ex}
                & 2$R_0$ & 0 & 0 & $-0.56$ (1.57) & $-0.61$ (1.54) & $-0.60$ (1.51) & $-0.04$ (0.77) & $-0.03$ (0.71) & \ \ 0.01 (0.65)\\
                & ... & 0 & 30 & $-0.57$ (1.57) & $-0.61$ (1.55) & $-0.61$ (1.53) & $-0.05$ (0.78) & $-0.03$ (0.74) & \ \ 0.01 (0.68)\\
                & ... & 45 & 30 & $-0.57$ (1.72) & $-0.61$ (1.68) & $-0.61$ (1.66) & $-0.04$ (0.85) & $-0.03$ (0.79) & \ \ 0.01 (0.72)\\
                & ... & 90 & 30 & $-0.58$ (1.88) & $-0.62$ (1.85) & $-0.61$ (1.82) & $-0.04$ (0.93) & $-0.04$ (0.86) & \ \ 0.00 (0.78)\\
                \rule{0pt}{3.5ex}
HE~0435$-$1123 & 0.5$R_0$ & 0 & 0 & \ \ 0.12 (0.42) & \ \ 0.00 (0.20) & \ \ 0.05 (0.31) & $-0.12$ (0.42) & $-0.07$ (0.48) & \ \ 0.05 (0.31) \\
                &  ... & 0 & 30 & \ \ 0.12 (0.43) & \ \ 0.00 (0.21) & \ \ 0.05 (0.31) & $-0.13$ (0.43) & $-0.07$ (0.49) & \ \ 0.05 (0.31)\\
                &  ... & 45 & 30 & \ \ 0.12 (0.47) & \ \ 0.00 (0.23) & \ \ 0.05 (0.35) & $-0.12$ (0.48) & $-0.07$ (0.54) & \ \ 0.05 (0.35)\\
                &  ... & 90 & 30 & \ \ 0.11 (0.52) & \ \ 0.00 (0.25) & \ \ 0.05 (0.39) & $-0.11$ (0.52) & $-0.07$ (0.60) & \ \ 0.05 (0.39)\\
                \rule{0pt}{3.5ex}
                & $R_0$ & 0 & 0 & \ \ 0.38 (1.06) & \ \ 0.00 (0.47) & \ \ 0.21 (0.80) & $-0.37$ (1.05) & $-0.17$ (1.24) & \ \ 0.20 (0.80)\\
                & ... & 0 & 30 & \ \ 0.38 (1.07) & \ \ 0.00 (0.49) & \ \ 0.20 (0.81) & $-0.38$ (1.07) & $-0.17$ (1.25) & \ \ 0.20 (0.82)\\
                & ... & 45 & 30 & \ \ 0.37 (1.16) & \ \ 0.00 (0.54) & \ \ 0.20 (0.91) & $-0.37$ (1.17) & $-0.17$ (1.38) & \ \ 0.19 (0.91)\\
                & ... & 90 & 30 & \ \ 0.36 (1.28) & \ \ 0.00 (0.59) & \ \ 0.18 (1.01) & $-0.35$ (1.28) & $-0.17$ (1.52) & \ \ 0.19 (1.02)\\
                \rule{0pt}{3.5ex}
                & 2$R_0$ & 0 & 0 & \ \ 0.83 (2.38) & \ \ 0.01 (1.08) & \ \ 0.65 (2.00) & $-0.83$ (2.38) & $-0.17$ (2.92) & \ \ 0.65 (2.01) \\
                & ... & 0 & 30 & \ \ 0.85 (2.42) & \ \ 0.00 (1.10) & \ \ 0.65 (2.03) & $-0.83$ (2.42) & $-0.19$ (2.96) & \ \ 0.65 (2.04)\\
                & ... & 45 & 30 & \ \ 0.85 (2.64) & \ \ 0.01 (1.21) & \ \ 0.64 (2.24) & $-0.83$ (2.64) & $-0.20$ (3.25) & \ \ 0.64 (2.25)\\ 
                & ... & 90 & 30 & \ \ 0.85 (2.87) & \ \ 0.00 (1.32) & \ \ 0.61 (2.48) & $-0.84$ (2.89) & $-0.23$ (3.57) & \ \ 0.62 (2.50)\\[0.4mm]
\hline
\end{tabular}
\label{tab:rmsdelay}
\end{table*}

As a final illustration of microlensing effect on time delays, we created examples of microlensed quasar light curves including these effects. We used the damped random walk (DRW) model, which has been shown to capture quasar variability relatively well \citep{Kelly2009,Kozlowski2010,MacLeod2010}, to generate the driving light curve $f(t)$. The driving light curve is modeled with a time scale $\tau$ = 90 days and a fractional variability of 15\%. This damping time $\tau$ is shorter than typical of quasars in order to make it easier to visualize the microlensing effects. The contribution from any point on the disk lags the driving light curve by $t_{lag} = (1+z_{s})(R - x \mbox{ sin }i)/c$, with a flux contribution of $f(t-t_{lag})G(\xi)$. This is then weighted either by a constant in the absence of microlensing or a magnification pattern when microlensing is present, resulting in a snapshot of the disk brightness. Repeating this as a function of time produces a set of evolving maps of the disk brightness, from which one can generate the observed light curve.

Figure \ref{fig:exmaplelc} shows four examples of light curves (LC1--LC4) spanning 120 days for image A of RXJ~1131$-$1231. The disk size is set to $R_0$. We show two examples for a face-on disk (LC1, LC2) and two examples for an inclined disk (LC3, LC4 with $PA=$ 45$^{\circ}$ and $i=$ 30$^{\circ}$), where the disk positions are indicated in Figure~\ref{fig:delay1131} for the latter two cases. The predicted mean shifts based on Equation~\ref{eqn:anal} are given in Table \ref{tab:javelin_lag}. If we shift the microlensed light curves by this delay, they match the input light curves with some small differences in structure due to the microlensing. It is important to note that these changes in the light curve structures are not due to any movement of the quasar relative to the magnification pattern, but are instead due to changes in the disk surface brightness with time that are differentially weighted by microlensing. 

We next used the AGN lag estimation algorithm \texttt{JAVELIN} \citep{Zu2011} to estimate the lags between the ``micro'' and ``nomicro'' light curves for the example light curves in Figure \ref{fig:exmaplelc}. We assumed a generic 5\% fractional uncertainty and treated the ``nomicro'' light curves as the driving light curve and fit for the lag of the ``micro'' light curves. The medians lags and their 68\% uncertainties are shown in Table \ref{tab:javelin_lag}. We see that the mean lags predicted by Equation~\ref{eqn:anal} agree well with the lags determined from the model light curves. 


\begin{table}
\small
\caption{\texttt{JAVELIN} lags of example light curves} 
\centering 
\begin{tabular}{d{2.0}d{0.0}d{0.3}}
\hline\hline 
\multicolumn{1}{c}{LC} &
\multicolumn{1}{c}{$\langle \delta t \rangle$ (days)} &
\multicolumn{1}{c}{\texttt{JAVELIN} (days)} \\
\hline
\rule{0pt}{2.5ex} 1 & 2.73 & 2.88^{+0.42}_{-0.43} \\[1.5mm]
2 & -1.32 & -1.22^{+0.39}_{-0.38} \\[1.5mm]
3 & 2.34 & 2.30^{+0.51}_{-0.47} \\[1.5mm]
4 & -1.14 & -1.07^{+0.37}_{-0.38} \\[0.5mm]
\hline
\end{tabular}
\label{tab:javelin_lag}
\end{table}

\section{Discussion and conclusion}
Using both simple models and full simulations, we have shown that microlensing leads to perturbations in time delays on the scale of the light crossing time of the quasar disk, on the order of $\sim$ days.
Although the accretion disk is held fixed in this work, these delays will vary with time as the observer, source, lens, and stars in the lens move relative to one another. The time scales for changes are summarized in \cite{Mosquera2011}. Because the optical depth to microlensing for lensed images is almost unity, these time delay perturbations should be present to varying degrees in all lensed quasars. 

We have been very conservative in illustrating the amplitude of the microlensing effect, scaling the disk size $R_0$ to a standard thin disk model with an Eddington ratio of $L/L_E=$ 0.1. Typical quasars probably have higher Eddington ratios (see, e.g., \citealt{Kollmeier2006}), while both microlensing  (e.g., \citealt{Morgan2010,Mosquera2011}) and ongoing continuum reverberation mapping studies (e.g., \citealt{Shappee2014, Fausnaugh2016}) find that the accretion disks are two to four times larger than predicted by the thin disk theory. Hence, even our $2R_0$ models may still be underestimating the overall effect. Particularly for face-on disks, the effect will also depend on the quasar variability model. 

A first consequence of this new effect is that the uncertainty in a lens time delay now has an additional contributor, $\sigma_\mu \sim$ days. Ignoring the lens galaxy, the two traditional sources of error are measurement error in the time delay ($\sigma_{\delta t}$) and fluctuations in the surface density along the line of sight ($\sigma_{los}$). Unless the microlensing delays can be determined, there is no need for time delay measurement errors that are 
significantly smaller, $\sigma_{\delta t} \ll \sigma_\mu$. This is analogous to two image lenses where there is no point in measuring delays significantly more accurately than the limit set by the contribution from the cosmic variance in the density along the line of sight, $\sigma_{\delta t } \ll \sigma_{los} \Delta t$. 

The microlensing delay effect is an absolute, rather than fractional, error. Therefore, it is more important for the short delays common in four-image lenses (due to the high degree of symmetry) such as the two lenses investigated here, while it will matter less for long delays. For example, the longest delay in HE~0435$-$1123 is approximately $\Delta t \simeq 14$~days, so a $\sigma_\mu=1$~day contribution from microlensing represents a 7\% floor to the utility of this lens for cosmology. The longest delay in RXJ~1131$-$1231 is $\Delta t \simeq 91$~days, and such a microlensing effect represents a fractional error of only 1\% that is comparable to the effects of large scale structure. Lenses with long time delays, which tend to be two image lenses, are therefore strongly favored for obtaining measurements of cosmological time delays. Unfortunately, two-image lenses supply fewer model constraints on the structure of the lens galaxy. \cite{Tagore2017} also showed that two image lenses will produce the most biased estimates of $H_0$ from time delays, while cruciform quads have the lowest biases. The presence of microlensing effects may well revise this conclusion, as cruciform lenses with short time delays will be strongly limited by the effects of microlensing delays. Since the microlensing delay is on the scale of the accretion disk size $R_0 \propto M_{BH}^{2/3} (L/L_E)^{1/3}$ (Equation~\ref{eqn:rdisk}), lenses with lower mass black holes and shorter disk light crossing times are also preferred.

The microlensing delays also affect searches for substructures in lens galaxies using time delay anomalies in lensed quasars, also known as millilensing. Depending on the properties of the subhalo, substructures are thought to introduce time delay perturbations on the order of fraction of a day \citep{Keeton2009}. The substructure perturbations can be investigated through measurements of time delay ratios, which are insensitive to line of light structures and less sensitive to degeneracies due to the radial mass profile of the primary lens. Particularly with the biases produced by the mean delay shifts seen for microlensing of the lamp post variability model, this new microlensing effect may make using delay anomalies to search for substructures problematic. 


In addition to being an important new systematic problem for time delay
cosmology, this effect may be a boon to quasar microlensing studies.  In theory, microlensing probes the size of the source ($R_0$ or equivalent), the mean mass of the microlensing stars $\langle M_* \rangle$, and the effective velocity $v_e$ of the source relative to the magnification patterns. The mean stellar mass determines the typical Einstein radius $R_E \propto \langle M_* \rangle^{1/2}$, and the effective velocity is a combination of the motion of
the observer, the microlensing stars and the peculiar velocities of
the lens and the source.  Unfortunately, the only observable with
physical units is the time scale of the microlensing variability 
$t_E \propto R_E/v_e$, which is a degenerate combination of the mean 
Einstein radius and the effective velocity. This means that one
must use prior estimates for one of the three variables of interest
($R_0$, $\langle M_*\rangle$ and $v_e$) in order to constrain the
other two.  Traditionally, we have constrained the peculiar velocities (e.g., \citealt{Poindexter2008,Dai2010,Morgan2010,Mosquera2013}) while most other groups have assumed a mean microlens mass.

The microlensing time delay effect provides a new observable with 
units that is directly related to the size of the emission region,
$R_0$.  Measuring the effect has the potential of eliminating the
need for (strong) priors in microlensing analyses.  For any 
particular image, the effect depends on the disk size and 
inclination, but the strong dependence of the effect on the
inclination of the disk and its orientation relative to the 
caustic networks created by the tangential magnification suggests 
that it may be possible to disentangle the two effects, particularly
for four image lenses.  The projected shape and orientation of the
disk is fixed on the sky, but the orientation of the caustic 
networks rotates from image to image, so the differences in the
microlensing time delay effects between the images should strongly
constrain both variables.  

\section*{Acknowledgments}

The authors thank J. Munoz, E. Mediavilla, P. Schechter and P. Schneider for discussions. CSK is supported by NSF grants AST-1515876 and AST-1515927. This research made use of Astropy, a community-developed core Python package for Astronomy (Astropy Collaboration, 2013).

\end{document}